\documentclass[prl,reprint,nofootinbib,notitlepage,floatfix,superscriptaddress]{revtex4-1}
\usepackage[pdftex]{graphicx}

\usepackage{amsmath,amssymb,amstext}
\usepackage{mathrsfs}

\usepackage{multirow}

\usepackage{pdfsync}

\usepackage{comment}
\usepackage{bbold}
\usepackage{dsfont}
\usepackage{color}
\usepackage{hyperref}

\newcommand{\ket}[1]{\left|#1\right>} 

\begin{document}
  \title{Ultrafast quantum interferometry with energy-time entangled photons}

  \author{Jean-Philippe W. MacLean}
  \email{jpmaclean@uwaterloo.ca}
  \affiliation{Institute for Quantum Computing, University of Waterloo, Waterloo,
  Ontario, Canada, N2L 3G1} 
  \affiliation{Department of Physics \& Astronomy, University of Waterloo,
  Waterloo, Ontario, Canada, N2L 3G1}
  \author{John M. Donohue}
  \email{john.matthew.donohue@upb.de}
  \affiliation{Institute for Quantum Computing, University of Waterloo, Waterloo,
  Ontario, Canada, N2L 3G1} 
  \affiliation{Department of Physics \& Astronomy, University of Waterloo,
  Waterloo, Ontario, Canada, N2L 3G1}
  \affiliation{Integrated Quantum Optics, Applied Physics, Paderborn University, 33098 Paderborn, Germany} 
  \author{Kevin J. Resch}
  \affiliation{Institute for Quantum Computing, University of Waterloo, Waterloo,
  Ontario, Canada, N2L 3G1} 
  \affiliation{Department of Physics \& Astronomy, University of Waterloo,
  Waterloo, Ontario, Canada, N2L 3G1}


  \begin{abstract}
    Many quantum advantages in metrology and communication arise from
    interferometric phenomena.  Such phenomena can occur on ultrafast time
    scales, particularly when energy-time entangled photons are employed.
    These have been relatively unexplored as their observation necessitates
    time resolution much shorter than conventional photon counters.
    Integrating nonlinear optical gating with conventional photon counters can
    overcome this limitation and enable subpicosecond time resolution.  Here,
    using this technique and a Franson interferometer, we demonstrate
    high-visibility quantum interference with two entangled photons, where the
    one- and two-photon coherence times are both subpicosecond.  We directly
    observe the spectral and temporal interference patterns, measure a
    visibility in the two-photon coincidence rate of $(85.3\pm0.4)\%$, and
    report a CHSH-Bell parameter of $2.42\pm 0.02$, violating the local-hidden
    variable bound by 21 standard deviations. The demonstration of energy-time
    entanglement with ultrafast interferometry provides opportunities for
    examining and exploiting entanglement in previously inaccessible regimes.
  \end{abstract}
   
  \maketitle 

 Interferometry based on entangled quantum states is essential for enhanced
 metrology and quantum communication. Quantum correlations can enable
 interferometric measurements with improved
 sensitivity~\cite{slussarenko_unconditional_2017} and
 resolution~\cite{mitchell_super-resolving_2004}, and quantum advantages have
 been found for interferometric applications involving optical coherence
 tomography~\cite{nasr_demonstration_2003}, precise measurements of optical
 properties~\cite{steinberg_dispersion_1992,kaiser_quantum_2017}, and even the
 detection of gravitational waves~\cite{aasi_enhanced_2013}.  In laser physics,
 the development of ultrafast light sources has led to innovations in atomic
 spectroscopy, time-resolved measurements for quantum chemistry, nonlinear
 optics, x-ray sources, with applications in health sciences and industrial
 machining~\cite{backus_high_1998}.  For quantum light, energy-time entangled
 photons can also be produced with temporal features on ultrafast time
 scales~\cite{dayan_two_2004, nasr_ultrabroadband_2008,
 sensarn_generation_2010} and the wide availability of pulsed lasers has made
 this regime accessible for quantum state
 engineering~\cite{grice_eliminating_2001,
 mosley_heralded_2008,donohue_spectrally_2016}.  However, quantum
 interferometry with these states is challenging as the interference time
 scales are below the resolution of standard photon
 detectors~\cite{hadfield_single-photon_2009, eisaman_invited_2011}.  To
 overcome detector limitations, optical techniques have been developed to
 directly observe energy-time entangled quantum states on ultrafast time
 scales~\cite{maclean_direct_2018,jin_experimental_2018} by building effective
 fast photon counters using ultrafast optical gating in conjunction with
 standard photon
 counters~\cite{kuzucu_time-resolved_2008,kuzucu_joint_2008,allgaier_fast_2017}.
 Extending the measurement of quantum interference to subpicosecond time scales
 will be essential for developing new applications with ultrafast energy-time
 states of light.

 An important class of interferometers that has been used to observe quantum
 interference effects with energy-time entangled photons was developed by
 Franson in 1989~\cite{franson_bell_1989}.  Photon pairs are sent through two
 unbalanced interferometers creating interference in the coincidence rate but
 not in the single-photon detection rates.  High-visibility interference was
 observed in such an interferometer using spontaneous parametric downconversion
 (SPDC)~\cite{kwiat_high-visibility_1993,peer_temporal_2005}.  Franson
 interferometers with energy-time entangled states have since become important
 for applications in long-distance quantum key
 distribution~\cite{tittel_violation_1998}, measuring entanglement in
 high-dimensional~\cite{thew_bell-type_2004} and multiphoton
 states~\cite{reimer_generation_2016,agne_observation_2017}, scaling quantum
 information tasks to larger
 dimensions~\cite{zhong_photon-efficient_2015,ikuta_four-dimensional_2018}, and
 improving molecular spectroscopy~\cite{raymer_entangled_2013}.
 When both the single-photon and two-photon coherence times are ultrafast,
 however, observing quantum interference effects with a Franson interferometer
 requires new techniques to overcome detector limitations and the original
 interferometer concept can be adapted to provide delays on shorter time
 scales. 

 In this work, we temporally resolve two-photon interference with subpicosecond
 timing resolution.  The detectors are implemented by optically gating the
 photons in a nonlinear medium via noncollinear sum-frequency generation (SFG)
 with a short gate
 pulse~\cite{kuzucu_time-resolved_2008,kuzucu_joint_2008,allgaier_fast_2017,maclean_direct_2018}.
 Using this technique and single-photon spectrometers, we measure both the
 joint temporal and joint spectral features of a spatially separated two-photon
 state at the output of a Franson interferometer. 

  \begin{figure}[b] \centering
    \includegraphics[scale=1.0]{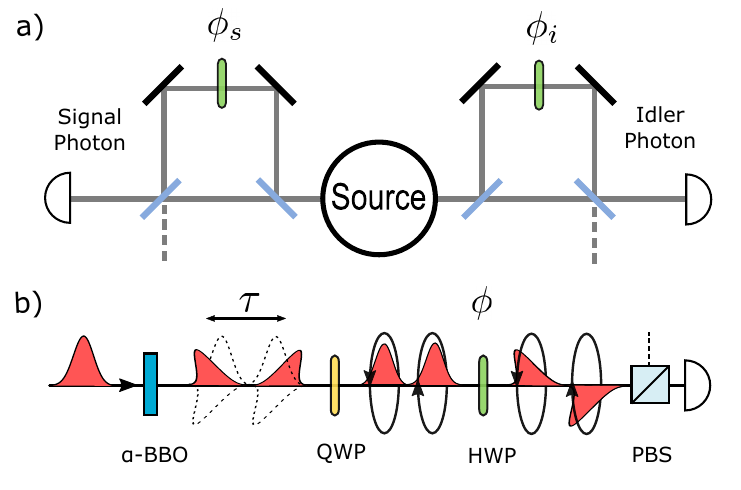}
    \caption{\footnotesize{Franson interferometer concept diagram.
    (a) Nonlocal interference can be seen by sending each photon of a
    two-photon energy-time entangled pair through an unbalanced interferometer.
    Each photon is split into early and late time bins and recombined with a
    phase applied to one bin.  (b) In each arm of the Franson interferometer, the
    delays and phases are implemented through birefringent material and
    wave-plates, creating a path difference on subpicosecond time scales between
    the short and long paths. A birefringent crystal ($\alpha$-BBO) splits a
    horizontally polarized photon into a diagonal and a delayed anti-diagonal
    mode. A quarter-wave plate (QWP) converts diagonal and anti-diagonal to
    left- and right-circularly polarized light.  A half-wave plate (HWP)
    introduces a phase between the circularly polarized photons. Both
    polarizations are then projected into the horizontal state with a
    polarizing beam splitter (PBS).}} 
    \label{fig:concept} 
  \end{figure}

 We produce energy-time entangled photon pairs with parametric downconversion
 pumped by a broadband laser pulse.  The photons are produced with strong
 anti-correlations between the signal, $\omega_s$, and idler, $\omega_i$,
 frequencies  leading to a narrow joint uncertainty $\Delta(\omega_s+\omega_i)$
 set by the bandwidth of the pump in broadly phasematched materials
 ~\cite{mikhailova_biphoton_2008, maclean_direct_2018}.  For a two-photon state
 with no spectral phase, the photon pairs will also exhibit strong correlations
 between the time of arrival of the signal, $t_s$, and the idler, $t_i$,
 leading to smaller joint uncertainty than their individual widths in time,
 $\Delta(t_s-t_i)<\Delta t_{s,i}$~\cite{maclean_direct_2018}.  A Franson
 interferometer, shown schematically in Fig.~\ref{fig:concept}(a), separates
 the photons on each side into a short and long path, with a time delay $\tau$,
 resulting in four possible combinations of paths.  The single-photon detection
 rates, which vary with the phase in each arm, $\phi_{s,i}$, have a coherence
 time inversely proportional to the single-photon spectral bandwidth,
 $\tau^{(1)}_{c_{s,i}}=1/\Delta\omega_{s,i}$, whereas the coincidence rate,
 which varies with $\phi_s+\phi_i$, has a two-photon coherence time inversely
 proportional to the two-photon spectral bandwidth,
 $\tau^{(2)}_c=1/\Delta(\omega_s+\omega_i)$ (see supplemental material).  When
 $\tau$ is set to be much larger than the single-photon coherence time but less
 than the two-photon coherence time, $\tau^{(1)}_{c_{s,i}}\ll\tau<
 \tau^{(2)}_c$, interference in the coincidence rate can be observed without
 any present in the single detection rates.  
 
  \begin{figure}[b!] \centering
    \includegraphics[scale=1]{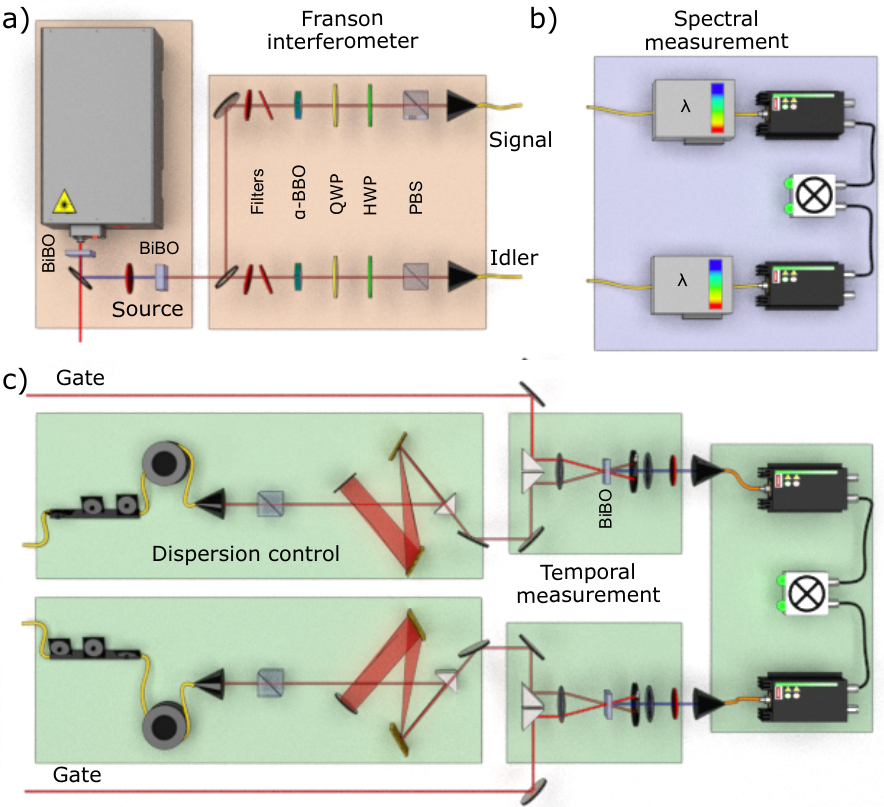}
    \caption{\footnotesize{Experimental setup.  (a)A Ti:sapphire laser pulse
      (775nm, 3.8W average power, 0.120~ps (s.d.) pulse-width), is frequency
      doubled in 2~mm of $\beta$-bismuth borate (BiBO).  After spectral
      filtering with a 0.2~nm FHWM bandpass filter, the second harmonic
      (387.6~nm, 300~mW average power, approximately 0.940~ps (s.d.) coherence
      time) pumps a 5~mm BiBO crystal for type-I spontaneous parametric
      downconversion (SPDC) generating frequency entangled photons centered at
      730~nm and 827~nm.  The photons are separated by a dichroic mirror and
      their bandwidth is controlled using tunable edge filters.  Each photon
      passes through an unbalanced interferometer consisting of $\alpha$-BBO,
      QWP, HWP, and PBS. We use 2.00~mm and 2.25~mm of $\alpha$-BBO to create a
      difference between the short and long paths of $\tau_s=0.820$~ps and
      $\tau_i=0.910$~ps on the signal and idler side, respectively.  The output
      of the Franson interferometer is coupled into single-mode fibers.  (b)
      Spectral measurements are made with single-photon spectrometers.  (c)
      Temporal measurements are performed using ultrafast gating with a strong
      laser pulse. A pair of grating compressors compensates for the dispersion
      introduced by the fibers.}} 
    \label{fig:setup} 
  \end{figure}

\begin{figure*}[t!]
     \centering
     \includegraphics[scale=1.0]{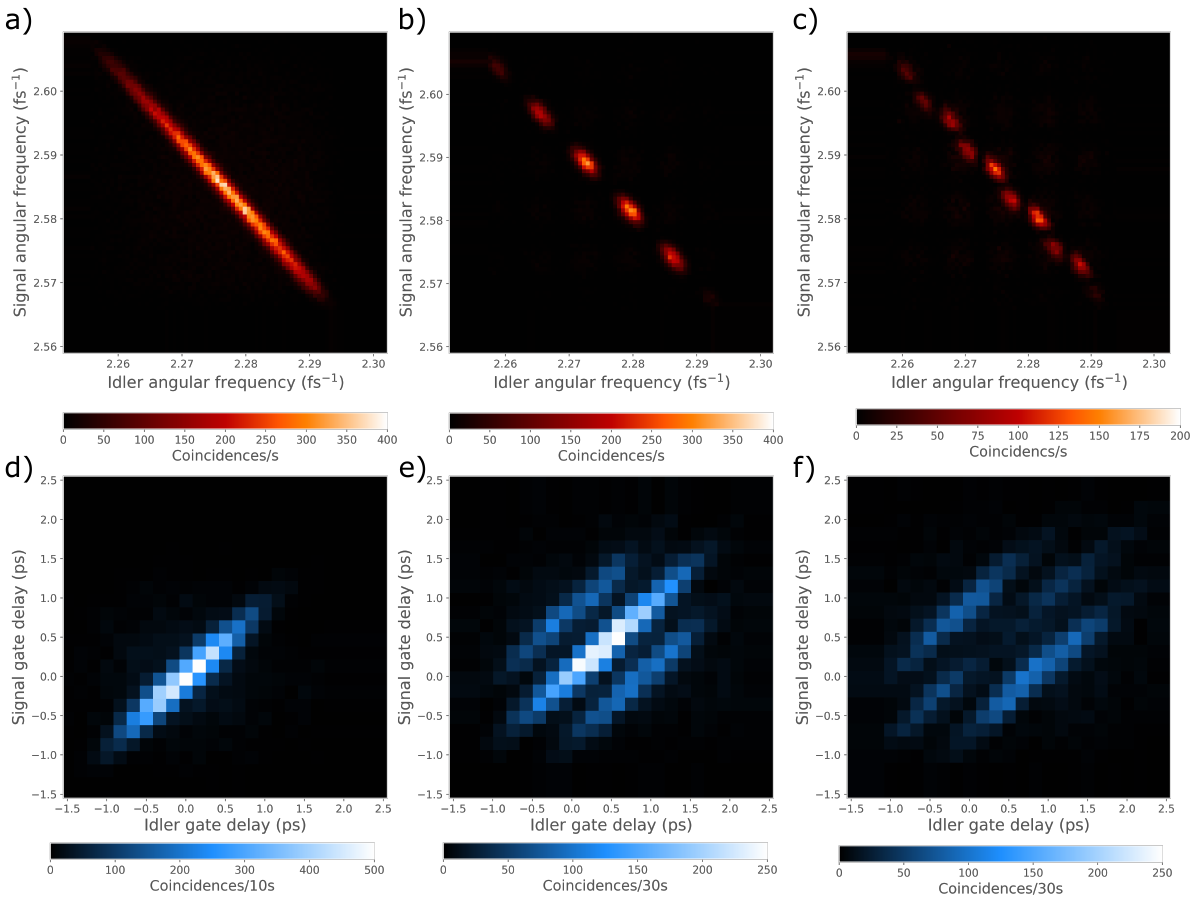} 
     \caption{\footnotesize{ 
     The  joint spectral intensity and joint temporal intensity of the
     two-photon state shown (a,d) before and (b,c,e,f) after the Franson
     interferometer.  After the interferometer, different fringe
     patterns are observed in the joint spectrum for (b) constructive and (c)
     destructive interference.  The interferometer shifts the temporal profile
     in (d) creating four different combinations of paths: short-short,
     short-long, long-short, and long-long.  We observe (e) constructive and
     (f) destructive interference in the central peak between the cases where
     the photons both take the short path and both take the long path. These
     correspond to two-photon states where the signal and idler phases sum to
   (b,e) $\phi_i+\phi_s=0$, and (c,f) $\phi_i+\phi_s=\pi$.  }}
     \label{fig:jointplots}
   \end{figure*}

 The interference in the coincidences results from the indistinguishability
 between the cases where the two photons both take the short path in the
 interferometer and where both take the long path. Meanwhile, the cases where
 they take opposite paths, labeled short-long and long-short, do not exhibit
 interference, thus limiting the visibility to 50\% without temporal
 resolution.  This, however, is the same maximum visibility that can be
 obtained in coincidence measurements with classically correlated light when
 zero visibility is observed in the single-photon rates~\cite{su_quantum_1991}.
 To observe higher visibility interference with energy-time entangled photons,
 sufficient time resolution is needed to resolve the arrival times of the early
 and late photons. This condition is typically met using continuous-wave-pumped
 downconversion sources whose two-photon coherence times are much longer.  They
 can therefore support interferometer delays in the range of 10~cm to
 1~m~\cite{brendel_pulsed_1999, grassani_micrometer-scale_2015,
 jaramillo-villegas_persistent_2017, peiris_franson_2017}, which can be
 implemented in free space or fiber, such that the time difference $\tau$
 between early and late photons (30~$\mu$s to 3~ns) remains much larger than
 standard detector resolution.

 We construct the ultrafast Franson interferometer using birefringent crystals
 where the long and short paths arise due to the different refractive indices,
 and hence different optical path lengths, for horizontally and vertically
 polarized light~\cite{donohue_coherent_2013}, as seen in
 Fig.~\ref{fig:concept}(b).  Two millimeters of $\alpha$-BBO creates an
 interferometer with relative delays below one picosecond and does not require
 any active phase stabilization.  The experimental setup is shown in
 Fig.~\ref{fig:setup}.  Signal-idler photon pairs are produced using SPDC with
 center wavelengths of 730~nm and 827~nm, respectively.  A pair of tunable edge
 filters in the source control the single-photon spectral bandwidths by making
 effective bandpass filters of 3.0~nm (s.d.) and 3.5~nm, for the signal and
 idler respectively.  The photon pairs are coupled into fiber, allowing for
 direct detection, spectral, or temporal measurements in coincidence, with or
 without the Franson interferometer. Spectral measurements are performed using
 two grating-based single-photon monochromators with a resolution of
 approximately 0.1~nm, while temporal measurements are implemented by optically
 gating the single-photons using SFG with femtosecond laser pulses which have
 an intensity pulse width of 0.120~ps (s.d.)~\cite{maclean_direct_2018}.

 The joint spectral intensity and joint temporal intensity of the state before
 the Franson interferometer were measured  and the data is shown in
 Figs.~\ref{fig:jointplots}(a) and \ref{fig:jointplots}(d), respectively.  We
 observe strong anti-correlations between the photon frequencies and strong
 positive correlations between their arrival times.  Measurements of the
 spectral widths in these plots allow us to estimate the one- and two-photon
 coherence times.  To account for the finite resolution of the spectrometers
 and temporal gates, Gaussian fits to the measured widths are deconvolved
 assuming Gaussian response functions.  The deconvolved frequency marginals
 (see supplemental material) are found to be $\Delta
 \omega_s=10.65~\textrm{ps}^{-1}$ and $\Delta \omega_i=9.57~\textrm{ps}^{-1}$,
 from which we estimate single-photon coherence times of
 $\tau^{(1)}_{c_s}=0.094$~ps and $\tau^{(1)}_{c_i}=0.105$~ps for the signal and
 idler respectively.  Gaussian fits to histograms of the spectral semi-minor
 and semi-major axes yield deconvolved two-photon spectral bandwidths of
 $\Delta(\omega_s+\omega_i)=1.531~\textrm{ps}^{-1}$ and
 $\Delta(\omega_s-\omega_i)=17.81~\textrm{ps}^{-1}$. From the former, we
 estimate a two-photon coherence time of $\tau^{(2)}_c=0.653$~ps.  The temporal
 measurements yield deconvolved temporal marginal widths of $\Delta
 t_s=0.455$~ps and $\Delta t_i=0.488$~ps and deconvolved temporal widths of the
 semi-minor and semi-major axes of $\Delta(t_s+t_i)=0.895$~ps and
 $\Delta(t_s-t_i)=0.091$~ps. 

   \begin{figure}[t!]
     \centering
     \includegraphics[scale=1.0]{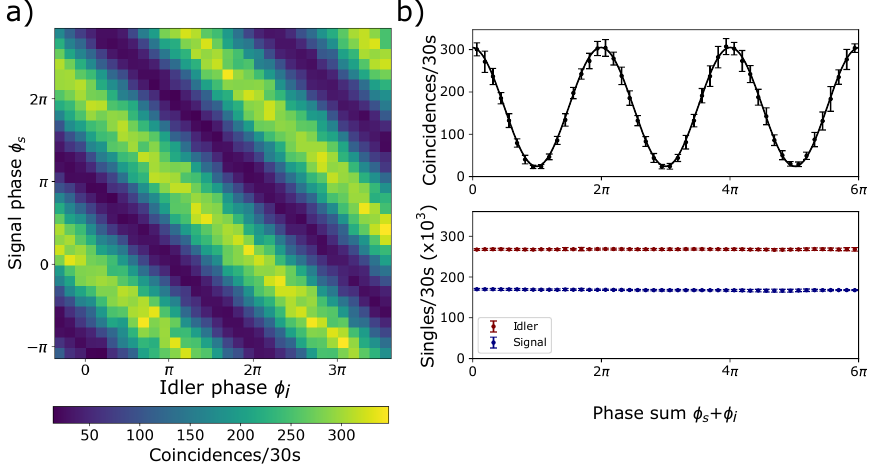} 
     \caption{\footnotesize{{Two-photon interference fringes.}}
     Franson interference between the upconverted signal and idler pair is
     measured by varying the signal and idler phases while setting the idler
     gate delay and signal gate delay halfway between the short and long paths
     of each side of the Franson interferometer. (a) We observe high-visibility
     interference with fringe oscillations along the diagonal which depend on
     the sum of the two phases $\phi_s+\phi_i$. (b) Weighted average of the
     coincidences and weighted average of the singles for the signal and idler
     pair as viewed as a function of their phase sum.  Interference fringes
     display oscillations of $(85.3\pm0.4)\%$ visibility, while the singles
     detection events show no apparent oscillations.
   }
     \label{fig:jointfransonphase}
   \end{figure}

 The joint spectral intensity and joint temporal intensity of the state
 \emph{after} the Franson interferometer are shown in
 Figs.~\ref{fig:jointplots}(b, c, e, f) for two different combinations of phase
 settings which provide the highest contrast between the constructive and
 destructive interference in the central peak of the temporal plots.  In
 Fig.~\ref{fig:jointplots}(b), we observe a joint spectral intensity similar to
 the one found in Fig.~\ref{fig:jointplots}(a) but with a periodic amplitude
 modulation.   The joint spectral intensity in Fig.~\ref{fig:jointplots}(c) is
 also modulated by two sinusoidal functions, but shifted with respect to the
 ones in Fig.~\ref{fig:jointplots}(b). These patterns correspond to the
 expected fringes for unbalanced interferometers applied to both the signal and
 idler, with phases $\phi_s+\phi_i=0$ and $\phi_s+\phi_i=\pi$, respectively. In
 the corresponding temporal plots, we observe constructive interference in
 Fig.~\ref{fig:jointplots}(e) and destructive interference in
 Fig.~\ref{fig:jointplots}(f), presenting, respectively, a strong peak and
 trough in the center of the distribution, while the two side peaks on either side
 of the central peak exhibit no interference.  Through these measurements, we
 are able to observe the effect of the interferometer in both spectral and
 temporal domains.

 We then measure the phase-dependent interference fringes of the Franson
 interferometer.  The signal and idler gate delays are set to 0.455~ps and
 0.410~ps, respectively, upconverting only the photons in the center of the
 joint temporal intensity where the highest contrast interference is observed,
 and corresponding to one pixel in Figs.~\ref{fig:jointplots}(e) and
 ~\ref{fig:jointplots}(f). The measured coincidences as a function of the
 signal and idler phases, $\phi_s,\phi_i$, are presented in
 Fig.~\ref{fig:jointfransonphase}.  We observe high-visibility interference
 fringes along one diagonal in Fig.~\ref{fig:jointfransonphase}(a) which
 corresponds to interference in the correlated phase setting, $\phi_s+\phi_i$,
 as expected for frequency anti-correlated photons (see supplemental material
 for further details).  From the same data set, we plot the integrated single
 and coincidence rates as a function of the phase sum, $\phi_s+\phi_i$, in
 Fig.~\ref{fig:jointfransonphase}(b). The coincidence rate exhibits fringes
 with $(85.4\pm0.4)\%$ visibility without background subtraction, whereas the
 single-photon rates exhibit no visible interference.  Error bars are obtained
 from a weighted average of the data points assuming Poissonian noise.

 To maximize the visibility of the Franson interference, we found that the
 interferometer delays need to shift the joint-temporal intensity in
 Fig.~\ref{fig:jointplots}(d) along its semi-major axis, such that a maximum
 overlap is obtained between the cases where both photons take the long path
 and where both photons take the short path. This can be achieved by matching
 the ratio of the applied interferometer delays $\tau$ to the ratio of the
 marginal temporal widths $\Delta t$, such that $\tau_i/\tau_s= \Delta
 t_i/\Delta t_s$.  The measured ratio was $\Delta t_i/\Delta t_s=1.07$,
 differing from unity due to the particular phase-matching conditions which can
 change the angle of the joint spectral amplitude
 function~\cite{grice_spectral_1997}.  We found that using different lengths of
 $\alpha$-BBO crystals, 2.00~mm  and 2.25~mm, created the appropriate temporal
 separations of approximately $\tau_s=0.820$~ps and $\tau_i=0.910$~ps, for the
 signal and idler, respectively, in order to approach this ratio and satisfy
 the conditions for two-photon interference.  We repeated the measurement when
 both crystal lengths were chosen to be 2.00~mm and observed a reduction of
 10\% in the visibility.

 The measured detector counts for each phase setting on the signal and idler
 sides can be viewed as one binary outcome ($+1$) of a projective measurement,
 where the corresponding outcome ($-1$) is obtained by a $\pi$ phase shift.
 Thus, we can look for a violation of the CHSH inequality from 16 combinations
 of signal-idler phases, 4 outcomes for each of the 4 joint projective
 measurements in the inequality~\cite{tittel_violation_1998} (see supplemental
 material).  We count for 200 seconds for each outcome, and obtain from these
 counts a CHSH-Bell parameter of $2.42\pm 0.02$, a violation of the
 local-hidden variable bound of 2 by 21 standard
 deviations~\cite{clauser_proposed_1969}. This is a consequence of the
 entanglement in our system and shows the high quality of the interference and
 the general performance of our measurement device.
 
 The visibility of the Franson interference and Bell violation could be further
 improved by reducing the second-harmonic generation (SHG) background from the
 laser in the optical gating.  From the measured upconversion rates after the
 source, we obtain a coincidence rate of about 44~Hz from which about 0.8~Hz
 can be attributed to the SHG background of the laser. This corresponds to a
 signal-to-noise ratio (SNR) of 54.  After the Franson interferometer, the
 measured coincidence rate at the peak is reduced by a factor of 4 but the SHG
 background remains the same, giving a SNR of 13.5. This translates to a
 reduction in visibility of 13\%, which accounts for most of the observed
 visibility loss.  The SHG background source could be reduced by utilizing a
 type-II process which would allow for additional polarization filtering.

 We have experimentally observed two-photon quantum interference on ultrafast
 time scales using a stable and compact Franson interferometer.  The optical
 gating detection mechanism enables the direct measurement of the joint
 temporal intensity as well as the observation of quantum interference
 phenomena and the violation of a CHSH-Bell inequality in a previously
 inaccessible regime.  In addition to interferometry, access to both spectral
 and temporal features will provide new tools for creating and characterizing
 two-photon states and will be essential for new applications in quantum state
 engineering, such as shaping ultrafast entangled photon pulses.

\begin{acknowledgments} 
 The authors would like to thank M. Mazurek and K. Fisher for fruitful
 discussions. This research was supported in part by the Natural Sciences and
 Engineering Research Council of Canada (NSERC), Canada Research Chairs,
 Industry Canada, and the Canada Foundation for Innovation (CFI).
\end{acknowledgments} 

\bibliography{ultrafast_quantum_interferometry}
 
\appendix 
\onecolumngrid
\clearpage
\section*{Supplemental Material} 

 The supplemental material is organized as follows. We first present additional
 fit parameters for the two-dimensional joint spectral intensity and joint
 temporal intensity in Figs.~\ref{fig:jointplots}(a) and
 ~\ref{fig:jointplots}(d), respectively. We next describe the experimental
 implementation of the unbalanced interferometers in Fig.~\ref{fig:concept} and
 Fig.~\ref{fig:setup}, and which were used to evaluate the CHSH inequality. We
 then consider the effect of finite correlations on the single-photon and
 two-photon detection rates at the output of the Franson interferometer. 

\section{Additional Experimental Details}

 Photons from the source were detected at a rate of 626,000 coincidence counts
 per second with $3.6\times10^6$ and $3.3\times10^6$ single-detection events
 per second for the signal and idler, respectively.  The heralded second-order
 coherence of the source, measured with a Hanbury Brown-Twiss interferometer,
 was $g^{(2)}(0)=0.391\pm0.004$ for the signal and $g^{(2)}(0)=0.395\pm0.006$
 for the idler. In general, double-pair emission will lead to a broad
 background in the joint spectrum and joint temporal intensity, however, due to
 the tight temporal filtering on both sides, we estimate that double pairs
 contribute to less than 1\% of the measured up-converted signal.  After the
 up-conversion on each side (without the Franson interferometer), approximately
 44 coincidence counts ($12,000$ up-converted signal singles and $21,000$
 up-converted idler singles per second) per second were measured at the peak,
 from which about 0.8 coincidence counts (400 signal and 2,500 idler singles)
 per second were background from the second harmonic of the gate pulse.

\begin{figure}[h] \centering
  \includegraphics{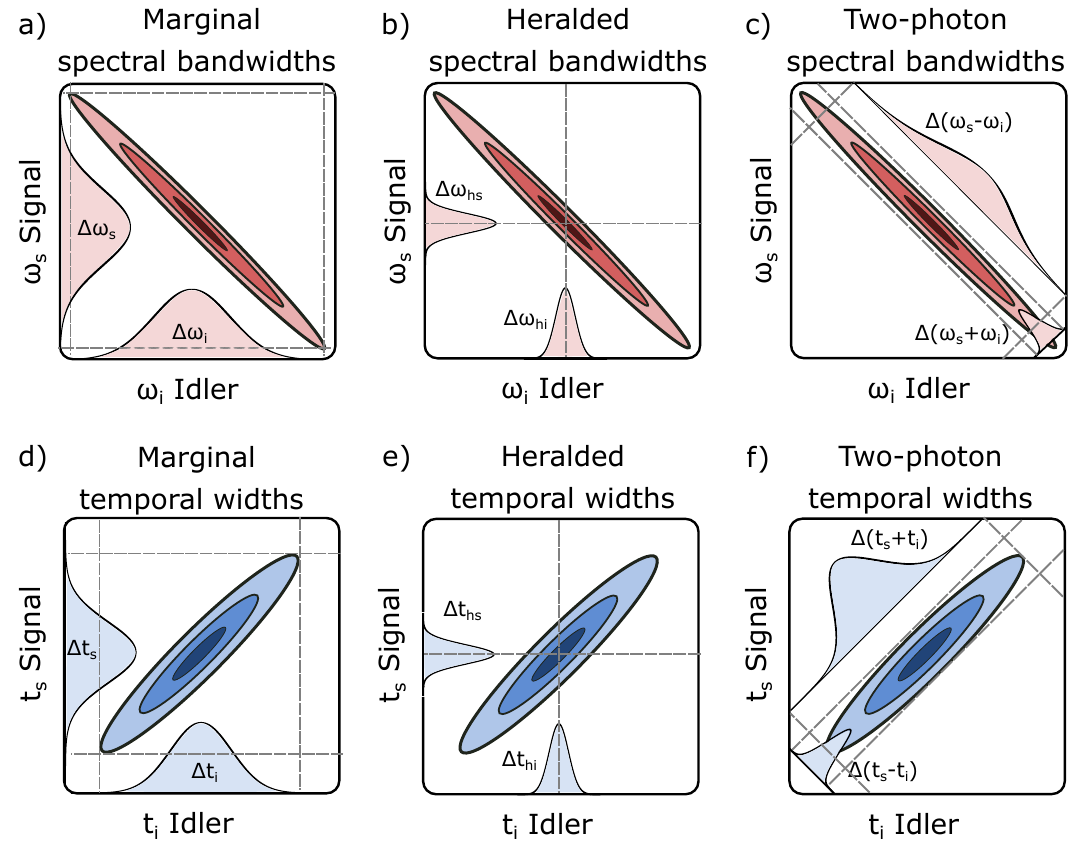} \caption{Spectral
    bandwidths and temporal widths for a frequency anti-correlated two-photon
    SPDC state. (a) The single-photon spectral bandwidth, $\Delta\omega$, is
    given by the marginal distribution obtained by projecting the joint
    spectral intensity onto either the signal or idler axes.  The single-photon
    coherence time, the time scale over which interference in the single-photon
    rates can occur, is related to the inverse of the single-photon spectral
    bandwidth, $\tau^{(1)}_{{c}}=1/\Delta\omega$. (b) The heralded spectral
    bandwidths, $\Delta\omega_h$, are the spectral bandwidths of the signal or
    idler photon when the frequency of the other is fixed. (c) The two-photon
    spectral bandwidths for the semi-minor, $\Delta(\omega_s+\omega_i)$, and
    semi-major axes, $\Delta(\omega_s-\omega_i)$, are obtained by projecting
    the joint spectral intensity along the corresponding diagonal axes
    $\omega_s\pm\omega_i$. The two-photon coherence time, the time scale over
    which interference in the coincidences can occur, is related to the inverse
    of the two-photon spectral bandwidth,
    $\tau^{(2)}_c=1/\Delta(\omega_s+\omega_i)$. (d-f) The marginal temporal
    widths, $\Delta t$, the heralded temporal widths, $\Delta t_h$, as well as
  the two-photon temporal widths, $\Delta(t_s\pm t_i)$ are obtained from the
joint temporal intensity in the same way as their spectral analogues.}
\label{fig:lengthscales} \end{figure}

 We present the fit parameters for the measured widths of the joint spectral
 intensity in Fig.~\ref{fig:jointplots}(a) and the joint temporal intensity in
 Fig.~\ref{fig:jointplots}(d) of the main text.  The marginal widths are
 obtained by fitting the marginals of Figs.~\ref{fig:jointplots}(a) and
 \ref{fig:jointplots}(d) to a one-dimensional Gaussian, while the heralded
 widths are obtained taking the average of several slices of the data when the
 frequency or time of one photon is fixed. A visual representation of the
 marginal and heralded widths is presented in Fig.~\ref{fig:lengthscales}.  The
 statistical correlation, $\rho$, is obtained by finding the value that best
 fits a two-dimensional Gaussian with the measured marginals.  The fit
 parameters are deconvolved assuming a Gaussian response function (see
 supplemental material of Ref.~\cite{maclean_direct_2018}), and these values
 are presented in parentheses alongside the values obtained from the raw
 measurements in Table~\ref{table:source-parameters}. 
 
\begin{table}[h]
  \centering
\caption{Fit parameters for the joint spectral intensity and the joint temporal
  intensity as seen in Fig.~\ref{fig:jointplots}(a) and \ref{fig:jointplots}(d)
  of the main text. All measured values are standard deviations and values in
  parentheses are deconvolved from a Gaussian response function.}
  \begin{tabular}{|c|c|c|c|}
    \hline
 \multicolumn{2}{|c|}{Property}	&  \multirow{2}{*}{Joint-spectrum} & Joint-temporal \\ 
 \multicolumn{2}{|c|}{(Deconvolved)} 	&				&intensity   \\ 
 \hline
 \multirow{5}{*}{Signal}& Center Frequency ($\omega$)	&$2584.6\pm0.4~\rm{ps}^{-1}$		& - 			\\\cline{2-4} 
 	&Marginal 				&$10.65\pm0.04~\rm{ps}^{-1}$ 		& $0.471\pm0.004$~ps 	\\ 
 	&width					& $(10.63\pm0.04~\rm{ps}^{-1}$) 	& $(0.455\pm0.004$~ps) 	\\ \cline{2-4}
 	&Heralded 				& $1.25\pm0.04~\rm{ps}^{-1}$ 		& $0.171\pm0.009$~ps    \\ 
	&width					& $(1.13\pm0.05~\rm{ps}^{-1})$ 		& $(0.059\pm0.022$~ps) 	\\\hline 
 \multirow{5}{*}{Idler}& Center Frequency  ($\omega$)	& $2276.7\pm0.4~ \rm{ps}^{-1}$   	& - 			\\\cline{2-4} 
 	&Marginal 				& $9.57\pm0.04~ \rm{ps}^{-1}$		& $0.502\pm0.005$~ps 	\\ 
 	&width					& $(9.56\pm0.04~\rm{ps}^{-1}$)		& $(0.488\pm0.005$~ps) 	\\\cline{2-4} 
 	&Heralded 				& $1.13\pm0.02~\rm{ps}^{-1}$     	& $0.183\pm0.010$~ps 	\\ 
 	&width					& $(1.02\pm0.02~\rm{ps}^{-1}$)    	& $(0.063\pm0.023$~ps) 	\\\cline{1-4}
\multicolumn{2}{|c|}{Statistical}	 	& $-0.9929\pm0.0001$			& $0.920\pm0.003$ 	\\ 
\multicolumn{2}{|c|}{Correlation}		& $(-0.9942\pm0.0001)$			& $(0.979\pm0.004)$ 	\\\hline 
\end{tabular}
\label{table:source-parameters}
\end{table}

\subsection{The unbalanced interferometers for ultrafast photons}

 The experimental implementation of the Franson interferometer presented in
 Fig.~\ref{fig:concept} was chosen to provide a stable and compact method of
 creating time bin states with subpicosecond temporal separations. In this
 section, we analyze the transformations applied to the polarization state of
 the photon by the unbalanced interferometer in Fig.~\ref{fig:concept}(b),
 composed of a birefringent crystal, wave plates, and a polarizing beam-splitter
 (PBS).  
 
 We denote the eigenstates of the Pauli operators $\sigma_z$ as
 $\ket{H}$ and $\ket{V}$, representing the horizontal and vertical polarization
 states of light.  After downconversion, the polarization state of each photon
 is vertical, $\ket{\psi}_{\textrm{pol}}=\ket{V}$. The $\alpha$-barium borate
 ($\alpha$-BBO) birefringent crystals at 45~degrees separate the photons on
 each side into early, $\ket{e}$, and late, $\ket{l}$, time bins with a
 temporal separation of $\tau_s=0.820$~ps and $\tau_i=0.910$~ps, for the signal
 and idler, respectively. The two time bins have orthogonal polarizations,
 which we denote as diagonal,
 $\ket{D}=\tfrac{1}{\sqrt{2}}\left(\ket{H}+\ket{V}\right)$ and anti-diagonal,
 $\ket{A}=\tfrac{1}{\sqrt{2}}\left(\ket{H}-\ket{V}\right)$. As a result, the
 polarization state is transformed to,
 $\ket{\psi}_{\textrm{pol}}\rightarrow\tfrac{1}{\sqrt{2}}\left(\ket{D}\ket{e}+\ket{A}\ket{l}\right)$.
 The phase difference $\phi$ between the two time bins can be controlled by
 manipulating the polarization of the two modes after the $\alpha$-BBO crystals
 with two wave plates and a PBS. A quarter-wave
 plate (QWP) first converts the two orthogonal polarization modes into
 left-circular, $\ket{L}=\frac{1}{\sqrt{2}}\left(\ket{H}-i\ket{V}\right)$, and
 right-circular polarizations,
 $\ket{R}=\frac{1}{\sqrt{2}}\left(\ket{H}+i\ket{V}\right)$, resulting in the
 state $\tfrac{1}{\sqrt{2}}\left(\ket{L}\ket{e}+\ket{R}\ket{l}\right)$.  A
 half-wave plate (HWP) at an angle $\theta$, described by the unitary operator, 
    \begin{align}
      U_{\textrm{HWP}}(\theta)=i  
      	\left(\begin{matrix}
	  \cos{2\theta} & \sin{2\theta} \\
	  \sin{2\theta} & -\cos{2\theta}
	\end{matrix}\right),
      \label{}
    \end{align}
 next applies the following transformations on the left- and right-circular
 polarizations of light,
 \begin{align}
      U_{\textrm{HWP}}(\theta)\ket{R}&=ie^{i2\theta}\ket{L}\\
      U_{\textrm{HWP}}(\theta)\ket{L}&=ie^{-i2\theta}\ket{R},
      \label{eq:HWP}
    \end{align}
 thus modifying the state to
 $\tfrac{1}{\sqrt{2}}i\left(e^{i2\theta}\ket{R}\ket{e}+e^{-i2\theta}\ket{L}\ket{l}\right)$.
 The PBS then erases the polarization information by projecting both circular
 polarizations into the horizontal mode $\ket{H}$, transforming the state to
 $\tfrac{1}{2}i\ket{H}\left(e^{-i2\theta}\ket{e}+e^{i2\theta}\ket{l}\right)$.
 As a result of these transformations, the photon at the output of the unbalanced
 interferometer is in a time bin state  with a phase difference between the
 early and late bins that can be set by the angle of the HWP through the
 parameterization $\phi=4\theta$. 

\subsection{Bell inequality using Franson interferometry}

 The measured detector counts for each phase setting $\phi$ in the unbalanced
 interferometers can be viewed as one binary outcome of a projective
 measurement, which we assign the value $(+1)$. The corresponding outcome
 ($-1$) could be obtained by placing a second detector to measure the photon
 events at the second output port of the unbalanced interferometer, however,
 here the second outcome $(-1)$ is instead obtained by measuring the photon
 events from the same detector but with an additional $\pi$ phase shift
 introduced in the interferometer using the HWPs in Fig.~\ref{fig:setup}.
 Given measurement outcomes $\pm1$ for two measurement choices labeled
 $a,a^\prime$ for the signal and $b,b^\prime$ for the idler, we measure the
 coincidence rates for the four outcomes of each joint projective measurement,
 denoted $R_{i,j}(a,b),(i,j=\pm1)$, and evaluate the correlation
 coefficient~\cite{tittel_violation_1998}, 
 \begin{align}
      E(a,b)=\frac{R_{++}(a,b)+R_{--}(a,b)-R_{+-}(a,b)-R_{-+}(a,b)}{R_{++}(a,b)+R_{--}(a,b)+R_{+-}(a,b)+R_{-+}(a,b)}.
      \label{eq:EBell}
    \end{align}
 Assuming a local-hidden variable model, the CHSH
 inequality~\cite{clauser_proposed_1969} provides an upper limit to the
 combination of four correlation coefficients, which can be written as,
    \begin{align}
      S=\left|E(a,b)+E(a,b^\prime)+E(a^\prime,b)-E(a^\prime,b^\prime)\right|\leq2.
      \label{eq:SBell}
    \end{align}
 In Table~\ref{table:Bell-counts}, we provide a table of raw coincidence counts
 for a particular combination of two projective measurements in the $x-z$ plane
 of the Bloch sphere on both the signal and idler sides.  From these counts, a
 CHSH Bell-parameter of $S=2.42\pm0.02$ is obtained, thus violating the
 inequality by 21 standard deviations.

\begin{table}[h]
  \centering
  \caption{Measured photon counts for CHSH-Bell inequality. The optical gate
    delays are set to upconvert photons in the center of the joint-temporal
    intensity. Upconverted coincidence counts are measured over 200 seconds for
    16 combinations of signal-idler phases in the Franson interferometer. These
    correspond to projective measurements performed on the signal and idler
    sides respectively labeled $a,a^\prime$ and $b,b^\prime$ with binary
    outcomes ($\pm1$) assigned for each phase setting. 
  }
 \begin{tabular}{|c|c|cc|cc|cc|}
   \hline
 \multicolumn{4}{|c}{}  & \multicolumn{4}{|c|}{Signal Phase $(\phi_s)$}  \\ \cline{5-8}
 \multicolumn{4}{|c|}{}	&$7\pi/4$ & $3\pi/4$ & $\pi/4$ & $5\pi/4$\\ \cline{5-8}
 \multicolumn{4}{|c}{}  & \multicolumn{2}{|c|}{$a$}&\multicolumn{2}{|c|}{$a^\prime$}  \\ 
 \multicolumn{4}{|c|}{}	&$(+1)$ & $(-1)$ & $(+1)$ & $(-1)$\\ \hline
	 		&$0$		&\multirow{2}{*}{$b$}		&$(+1)$	& 1292	& 367	& 1419 	& 336  	\\
 Idler Phase	 	&$\pi$		&				&$(-1)$ & 315 	& 1331 	& 329	& 1394	\\\cline{2-8}
 $(\phi_i)$		&$\pi/2$	&\multirow{2}{*}{$b^\prime$}	&$(+1)$ & 1423 	& 294 	& 358	& 1333	\\
 			&$3\pi/2$	&				&$(-1)$	& 301 	& 1469 	& 1401	& 335	\\
 \hline
\end{tabular}
\label{table:Bell-counts}
\end{table}

\section{Franson interferometry with finite correlations}

 In this section, we first calculate the overall coincident and single-photon
 detection rates of an energy-time entangled two-photon state after the Franson
 interferometer. We will show that this leads to two distinct time scales of
 interference for the single-photon detection rates and the coincidence
 detection rate. We then describe the need for temporal selection to improve
 the visibility of the interference in the coincidence rate after the Franson
 interferometer, and describe the effect of spectral or temporal selection by
 calculating the joint spectrum and joint temporal intensity for two-photon
 state after the Franson interferometer.  Finally, we discuss the parameters
 for optimizing the visibility of the two-photon interference.  
 
 Consider the two-mode state with signal $\omega_s$ and idler $\omega_i$
 frequency modes,  
 \begin{align}
     \ket{\psi}=\int
     d\omega_sd\omega_iF(\omega_s,\omega_i)
     a^{\dagger}_{\omega_s}a^{\dagger}_{\omega_i}\ket{0}.
     \label{eq:SPDC_state}
   \end{align}
 At the source, the joint spectral amplitude $F(\omega_s,\omega_i)$ of a
 pure two-mode state with no spectral phase can be described in Gaussian form as, 
 \begin{align}
   \begin{split}
   &F_{\textrm{source}}\left( \omega_s,\omega_i \right) = \\
   &\frac{1}{\sqrt{2\pi\sigma_{\omega_s}\sigma_{\omega_i} } \left( 1 - \rho_{\omega}^2 \right)^{1/4}} 
   \exp \left( { -\frac{1}{{2\left( {1 - {\rho_{\omega} ^2}} \right)}}\left[ {\frac{{{{\left(
     {{\omega _s} - {\omega _{s0}}} \right)}^2}}}{{2{\rm{\sigma }}_{\omega_s}^2}} 
     + \frac{{{{\left( {{\omega _i} - {\omega _{i0}}} \right)}^2}}}{{2\sigma
       _{\omega_i}^2}} 
 - \frac{\rho_{\omega} \left(\omega_s-\omega_{s0}\right)\left(\omega _i-\omega
 _{i0}\right)}{\sigma_{\omega_s}\sigma_{\omega_i}}} \right]}
 \right),
 \end{split}
 \label{eq:JSA}
  \end{align}
 where $\sigma_{\omega_s}$ and $\sigma_{\omega_i}$ are the marginal bandwidths
 of the signal and idler, respectively, and where the correlation parameter
 $\rho_{\omega}=\Delta(\omega_s\omega_i)/\Delta \omega_s \Delta \omega_i$
 describes the statistical correlations between the frequency of the signal and
 idler modes, and can be related to the purity of the partial trace,
 $P=\sqrt{1-\rho_{\omega}^2}$. For frequency anti-correlated photons, as shown
 in Fig.~\ref{fig:jointplots}(a), the frequency correlations are negative and
 $\rho_{\omega}<0$.

The Franson interferometer introduces delays $\tau_s$ and $\tau_i$ between the
short and long arms of each unbalanced interferometer with a phase $\phi_s$ and
$\phi_i$ on the signal and idler sides, respectively.  The joint spectral
amplitude after the Franson interferometer takes the form,
  \begin{align}
    F_{\textrm{franson}}\left( \omega_s,\omega_i \right)=
    F_{\textrm{source}}\left( \omega_s,\omega_i \right)
   \times\frac{1}{4}\left(1+e^{i(\omega_s\tau_s+\phi_s)}\right)
   \left(1+e^{i(\omega_i\tau_i+\phi_i)}\right). 
    \label{eq:JSAF}
  \end{align}
 The overall coincidence rate directly after the interferometer is,
 \begin{align}
    \begin{split}
      C\left( {{\phi _s},{\phi _i}} \right) &=\int_{-\infty}^{\infty}
      d\omega_sd\omega_i\left|F_{\rm{franson}}(\omega_s,\omega_i)\right|^2 \\
    &\propto 1 + \exp \left( {\frac{{ - \tau
      _s^2\sigma_{\omega_s}^2}}{2}} \right)\cos \left( {{\omega _{s0}}{{\rm{\tau }}_s} -
      {\phi _s}} \right) + \exp \left( {\frac{{ - \tau _i^2\sigma_{\omega_i}^2}}{2}}\right)\cos \left( {{\omega
	  _{i0}}{{\rm{\tau }}_i} - {\phi _i}} \right) \\
	  &+ \frac{1}{2}\exp \left( { - \frac{1}{2}{{\left( {{\sigma_{\omega_s}}{{\rm{\tau
	  }}_s} - {\sigma_{\omega_i}}{{\rm{\tau }}_i}} \right)}^2} - \left( {1 + \rho_{\omega} }
	  \right){\sigma_{\omega_s}}{\sigma_{\omega_i}}{{\rm{\tau }}_s}{\tau _i}} \right)\cos \left[
    {\left( {{\omega _{s0}}{\tau _s} - {\phi _s}} \right) + \left( {{\omega
      _{i0}}{\tau _i} - {\phi _i}} \right)} \right] \\
      &+ \frac{1}{2}\exp \left( { - \frac{1}{2}{{\left( {{\sigma_{\omega_s}}{{\rm{\tau
      }}_s} - {\sigma_{\omega_i}}{{\rm{\tau }}_i}} \right)}^2} - \left( {1 - \rho_{\omega} }
      \right){\sigma_{\omega_s}}{\sigma_{\omega_i}}{{\rm{\tau }}_s}{\tau _i}} \right)\cos \left[
   {\left( {\omega _{s0}\tau _s - {\phi _s}} \right) - \left( {{\omega
     _{i0}}{\tau _i} - {\phi _i}} \right)} \right].
	\end{split}
	  \label{eq:CCFranson}
  \end{align}
 For frequency anti-correlated photons, $\rho\rightarrow-1$, we expect
 interference which depends on the phase sum $\phi_s+\phi_i$, whereas for
 frequency correlated photons $\rho\rightarrow1$, the interference depends on
 the phase difference $\phi_s-\phi_i$. Considering the idealized case of
 frequency anti-correlations ($\rho\rightarrow-1$), assuming the signal and
 idler photon bandwidths $\sigma_{\omega}$ are the same, and the interferometer
 delays $\tau$ are equal, Eq.~\ref{eq:CCFranson} simplifies to, 
  \begin{align}
    \begin{split}
      C\left( {{\phi _s},{\phi _i}} \right) 
    &\propto 1 + \exp \left( {\frac{{ - \tau^2\sigma_{\omega}^2}}{2}} \right)\cos \left( {{\omega _{s0}}{{\rm{\tau}}} -
      {\phi _s}} \right) + \exp \left( {\frac{{ - \tau^2\sigma_{\omega}^2}}{2}}\right)\cos \left( {{\omega _{i0}}{{\rm{\tau}}} - {\phi _i}} \right) \\
    &+\frac{1}{2}\exp{\left(- \left( 1 + \rho_{\omega}
    \right){\sigma_{\omega}^2}\tau^2\right)} \cos \left[
    {\left( {{\omega _{s0}}{\tau} - {\phi _s}} \right) + \left( {{\omega
      _{i0}}{\tau} - {\phi _i}} \right)} \right]. 
	\end{split}
	  \label{eq:CCFranson2}
  \end{align}
 On the other hand, single-photon detection events have interference fringes
 described by, 
  \begin{align}
    S(\phi_j)\propto
    1+\exp\left(-\frac{1}{2}\sigma_{\omega_j}^2\tau_j^2\right)\cos\left(\omega_{j0}\tau_j-\phi_j\right).
    \label{eq:SFranson}
  \end{align}
 where $j\in\{s,i\}$. Comparing Eq.~\ref{eq:CCFranson2} and
 Eq.~\ref{eq:SFranson}, we find there are two time scales for interference for
 the two-photon state from downconversion.  The single-photon interference in
 Eq.~\ref{eq:SFranson} varies with $\phi_j$ and has a coherence time that
 depends on the inverse bandwidth of the photons,
 $\tau_c^{(1)}=1/\sigma_{\omega}=1/\Delta\omega$, whereas the two-photon
 interference in Eq.~\ref{eq:CCFranson2} varies with the sum $\phi_s+\phi_i$
 and has a coherence time that depends on the two-photon spectral bandwidth,
 $\tau_c^{(2)}=1/\left(\sqrt{2}\sqrt{1+\rho}\sigma_{\omega}\right)=1/\Delta(\omega_s+\omega_i)$.
 The Franson interferometer can thus be used to separate these two time scales
 by setting the delay $1/\Delta\omega\ll\tau\le1/\Delta(\omega_s+\omega_i)$
 between the single-photon and two-photon coherence times. Thus, with the
 appropriately chosen delay settings, we find the singles detection rates are
 constant whereas the coincident detection rate has oscillating fringes which
 depends on $\phi_s+\phi_i$ with an interference visibility of
 $V=\frac{1}{2}\exp{\left(- \frac{1}{2}\Delta(\omega_s+\omega_i)^2\tau^2\right)}$. 
 The visibility
 $V\le\frac{1}{2}$ without temporal selection is limited by the
 non-interfering background contributions from the short-long and long-short
 paths of the interferometer. In order to improve the measured visibility,
 these non-interfering background terms must be temporally filtered.

\subsubsection{Franson interferometry with spectral or temporal selection}

 We now discuss the spectral and temporal features of the downconverted state
 after the Franson interferometer. This is achieved by calculating the joint
 spectrum and joint temporal intensity.  The joint spectrum is obtained from
 the modulus squared of Eq.~\ref{eq:JSAF},
  \begin{align}
    \left|F_{\textrm{franson}}\left( \omega_s,\omega_i \right)\right|^2= 
    \left|F_{\textrm{source}}\left( \omega_s,\omega_i \right)\right|^2
    \cos\left(\frac{\omega_{s0}\tau_s+\phi_s}{2}\right)^2
    \cos\left(\frac{\omega_{i0}\tau_i+\phi_i}{2}\right)^2.
    \label{eq:JSI_Franson}
  \end{align}
 It consists of the original source spectrum
 $\left|F_{\rm{source}}(\omega_s,\omega_i)\right|^2$, which is intensity
 modulated. When  $\phi_i+\phi_s=0$, the oscillations for the anti-correlated
 frequencies remain in phase, as in Fig.~\ref{fig:jointplots}(b) of the main
 paper, whereas when $\phi_s+\phi_i=\pi$, they will be out of phase, as in
 Fig.~\ref{fig:jointplots}(c).  
 
 The joint temporal amplitude is obtained by
 taking the Fourier transform of the joint spectral amplitude,
 \begin{align}
   f_{\textrm{franson}}\left( {{t_s},{t_i}} \right) =
   \int d{\omega _i}d{\omega _s}F_{\textrm{franson}}\left( {{{\rm{\omega }}_i},{{\rm{\omega }}_s}}
   \right){e^{i{\omega _i}{t_i}}}{e^{i{\omega _s}{t_s}}}
 \end{align}
 from which we can obtain the joint temporal intensity,
 \begin{align}
   \begin{split}
  \left|f_{\textrm{franson}}\left( {{t_s},{t_i}} \right)\right|^2&\propto
  f_{ss}^2\left(t_s,t_i\right)+f_{ls}^2\left(t_s,t_i\right)+f_{sl}^2\left(t_s,t_i\right) +f_{ll}^2\left(t_s,t_i\right)\\
  &+
  2\left[f_{ss}\left(t_s,t_i\right)f_{ls}\left(t_s,t_i\right)+f_{sl}\left(t_s,t_i\right)f_{ll}\left(t_s,t_i\right)\right]\cos
  \left( {{\omega _{s0}}{{\rm{\tau }}_s} - {\phi _s}} \right)\\
  &+ 2\left[f_{ss}\left(t_s,t_i\right)f_{sl}\left(t_s,t_i\right)+f_{ls}\left(t_s,t_i\right)f_{ll}\left(t_s,t_i\right)\right]\cos \left( {{\omega _{i0}}{{\rm{\tau }}_i} - {\phi _i}} \right) \\
  &+ 2f_{sl}\left(t_s,t_i\right)f_{ls}\left(t_s,t_i\right)\cos \left[ {\left( {\omega _{s0}\tau _s - {\phi _s}}
      \right) - \left( {{\omega _{i0}}{\tau _i} - {\phi _i}} \right)} \right]\\
  &+ 2f_{ss}\left(t_s,t_i\right)f_{ll}\left(t_s,t_i\right)\cos \left[ {\left( {{\omega _{s0}}{\tau _s} - {\phi _s}} \right) + \left( {{\omega
      _{i0}}{\tau _i} - {\phi _i}} \right)} \right], 
  \end{split}
  \label{eq:JTI_Franson}
 \end{align}
 where
 \begin{align}
   f_{ss}\left(t_s,t_i\right)&=\exp\left[-\left(\sigma_{\omega_s}t_s-\sigma_{\omega_i}t_i\right)^2-2(1+\rho)\sigma_{\omega_s}\sigma_{\omega_i}t_st_i\right]\\
   f_{ls}\left(t_s,t_i\right)&=\exp\left[-\left(\sigma_{\omega_s}(t_s+\tau_s)-\sigma_{\omega_i}t_i\right)^2-2(1+\rho)\sigma_{\omega_s}\sigma_{\omega_i}(t_s+\tau_s)t_i\right]\\
   f_{sl}\left(t_s,t_i\right)&=\exp\left[-\left(\sigma_{\omega_s}t_s-\sigma_{\omega_i}(t_i+\tau_i)\right)^2-2(1+\rho)\sigma_{\omega_s}\sigma_{\omega_i}t_s(t_i+\tau_i)\right]\\
   f_{ll}\left(t_s,t_i\right)&=\exp\left[-\left(\sigma_{\omega_s}(t_s+\tau_s)-\sigma_{\omega_i}(t_i+\tau_i)\right)^2-2(1+\rho)\sigma_{\omega_s}\sigma_{\omega_i}(t_s+\tau_s)(t_i+\tau_i)\right]
 \end{align}
 are the four terms that represent the different combinations of paths the
 photons can take in the Franson interferometer, either short-short ($f_{ss}$),
 long-short ($f_{ls}$), short-long ($f_{sl}$), or long-long ($f_{ll}$).  These
 are two-dimensional correlated Gaussian functions that are shifted with
 respect to the origin by the applied delays $\tau_i$ and $\tau_s$.  Different
 types of interference can occur between these paths. The first line in
 Eq.~\ref{eq:JTI_Franson} contains the non-interference terms, the second and
 third lines accounts for single-photon interference, while the fourth and
 fifth lines account for nonlocal two-photon interference, which depends on the
 overlap between $f_{ls}$ and $f_{sl}$ and between $f_{ss}$ and $f_{ll}$,
 respectively. For anti-correlated photons ($\rho\rightarrow$-1), the
 short-long $f_{sl}$ and long-short $f_{ls}$ terms do not overlap and the
 fourth line goes to zero since $f_{sl}f_{ls}\rightarrow0$.  The single-photon
 temporal marginal, on the other hand, is given by,
 \begin{align}
   \begin{split}
   \left|f_{\textrm{marginal}}(t)\right|^2&\propto\exp \left( { - 2{t^2}\left( {1 - {\rho _\omega }^2} \right){\sigma _\omega
   }^2} \right) 
   + \exp \left( { - 2{{\left( {t + \tau } \right)}^2}\left( {1 - {\rho _\omega
   }^2} \right){\sigma _\omega }^2} \right) \\
   &+ 2\exp \left( { - 2{t^2}\left( {1- {\rho _\omega }^2} \right){\sigma
   _\omega }^2{{\left( {t + \frac{\tau }{2}} \right)}^2} - \frac{1}{2}{\sigma
     _\omega }^2{\tau _\omega }^2} \right)\cos \left( {\tau {\omega _0} - \phi
   } \right). 
   \end{split}
   \label{eq:TI_Franson}
 \end{align}
 Comparing Eq.~\ref{eq:JTI_Franson} and Eq.~\ref{eq:TI_Franson}, we find, as
 before, two different timescale for two-photon and single-photon interference.
 The interference term which varies as $\phi_s+\phi_i$ depends on the overlap
 between $f_{ss}$ and $f_{ll}$, whereas the single-photon interference has a
 coherence time that depends on the inverse bandwidth
 ($1/\sigma_{\omega}=1/\Delta\omega$) of the downconverted light. 

 In order to calculate the expected coincidence and single-photon rates with
 temporal selection, we consider the limiting case where we temporally select
 only the photon arrival times halfway between the short and long paths. This
 is equivalent to setting $t_s=-\tau_s/2$ and $t_i=-\tau_i/2$ in
 Eq.~\ref{eq:JTI_Franson} and Eq.~\ref{eq:TI_Franson}, which simplify to,
 \begin{align}
   \label{eq:JTI_tau}
   \begin{split}
   \left|f_{\textrm{franson}}\left(-\frac{\tau_s}{2},-\frac{\tau_i}{2}\right)\right|^2&\propto
   2\exp\left( { - \frac{1}{2}\sigma_{\omega_s}^2\tau _s^2 -
   \frac{1}{2}{\sigma_{\omega_i}^2}\tau _i^2} \right)\left[\cos\left( {{\omega
     _{s0}}{\tau _s} - {\phi _s}} \right) + \cos\left( {{\omega _{i0}}{\tau _i}
     - {\phi _i}} \right)\right] \\
     &+ \exp\left( { - \frac{1}{2}{{\left( {{\sigma_{\omega_s}}{\tau _s} -
     {\sigma_{\omega_i}}{\tau _i}} \right)}^2} - \left( {1 + {\rho
     _\omega }} \right){\sigma_{\omega_s}}{\sigma_{\omega_i}}{\tau _s}{\tau _i}}
     \right)\left[1+\cos\left( {{\omega _{s0}}{\tau _s} + {\omega _{i0}}{\tau _i} -
     {\phi _s} - {\phi _i}} \right)\right] 
   \end{split}\\
   \left|f_{\textrm{marginal}}\left(-\frac{\tau}{2}\right)\right|^2&\propto\exp \left( { - \frac{1}{2}\left( {1 - {\rho _\omega }^2} \right){\sigma
   _\omega }^2{\tau ^2}} \right) + \exp \left( { - \frac{1}{2}{\sigma _\omega
   }^2{\tau ^2}} \right)\cos \left( {{\omega _0}\tau - \phi } \right)
   \label{}
 \end{align}
 The visibility of the two-photon interference term in Eq.\ref{eq:JTI_tau} is
 then maximized under two conditions: the ratio of the delays is proportional
 to the ratio of the marginal bandwidths,
 $\sigma_{\omega_s}\tau_s=\sigma_{\omega_i}\tau_i$, and the 
 delays are less than the two-photon coherence time,
 $\tau_s\tau_i\ll1/\left[2(1+\rho)\sigma_{\omega_s}\sigma_{\omega_i}\right]$.  Under these
 conditions, assuming the photon bandwidths are equal, and substituting the
 expressions for the single- and two-photon spectral bandwidths, the
 coincidence rate and single-photon rates of photon detections with temporal
 selection at $t_i=-\tau_i/2$ and $t_s=-\tau_s/2$ become,
 \begin{align}
   \begin{split}
   C(\phi_s,\phi_i)&=\left|f_{\textrm{franson}}\left(-\frac{\tau_s}{2},-\frac{\tau_i}{2}\right)\right|^2\propto
   \exp\left( { -\frac{1}{2}\Delta\left(\omega_s+\omega_i\right)^2\tau^2 }
   \right)\left[1+\cos\left({{\omega _{s0}}{\tau _s} + {\omega _{i0}}{\tau _i} -
     {\phi _s} - {\phi _i}}\right)\right] 
   \end{split}\\
   S(\phi_j)&=\left|f_{\textrm{marginal}}\left(-\frac{\tau}{2}\right)\right|^2\propto
   \exp \left( { - \frac{1}{2}\left( {1 - {\rho _\omega }^2}
   \right){\Delta\omega_j}^2{\tau ^2}} \right) + \exp \left( {
     -\frac{1}{2}\Delta\omega_j^2{\tau ^2}} \right)\cos \left( \omega_0\tau-\phi_j\right)
   \label{}
 \end{align}
 As before, single-photon interference is removed by making the delays larger
 than the single-photon coherence time, $\tau\gg1/\Delta\omega$. However now,
 with temporal selection, the non-interfering terms have been filtered and
 100\% interference visibility can be achieved in the two-photon coincidence
 rate.

\end{document}